\documentclass{article} 
\usepackage{iclr2027_conference,times}

\usepackage{amsmath,amsfonts,bm}

\def\eqref#1{equation~\ref{#1}}

\def\1{\bm{1}}

\DeclareMathAlphabet{\mathsfit}{\encodingdefault}{\sfdefault}{m}{sl}
\SetMathAlphabet{\mathsfit}{bold}{\encodingdefault}{\sfdefault}{bx}{n}

\usepackage{hyperref}
\usepackage{url}

\usepackage{booktabs}
\usepackage{graphicx}
\usepackage{multirow}
\usepackage[section]{placeins}
\usepackage{hyperref}
\usepackage{url}
\usepackage{xcolor}
\usepackage{listings}

\definecolor{promptblue}{RGB}{43,92,157}
\definecolor{promptgreen}{RGB}{61,130,73}
\definecolor{promptgray}{RGB}{248,249,251}
\definecolor{prompttext}{RGB}{35,38,42}
\lstdefinestyle{casecode}{
  basicstyle=\ttfamily\scriptsize,
  backgroundcolor=\color{promptgray},
  frame=single,
  rulecolor=\color{promptblue},
  framerule=0.8pt,
  framesep=5pt,
  xleftmargin=4pt,
  xrightmargin=4pt,
  breaklines=true,
  breakatwhitespace=false,
  columns=fullflexible,
  keepspaces=true,
  showstringspaces=false,
  aboveskip=5pt,
  belowskip=6pt
}
\lstdefinestyle{prompt}{
  basicstyle=\ttfamily\fontsize{6.4}{7.1}\selectfont\color{prompttext},
  backgroundcolor=\color{promptgray},
  frame=single,
  rulecolor=\color{promptgreen},
  framerule=0.8pt,
  framesep=5pt,
  xleftmargin=4pt,
  xrightmargin=4pt,
  breaklines=true,
  breakatwhitespace=false,
  columns=fullflexible,
  keepspaces=true,
  showstringspaces=false,
  aboveskip=5pt,
  belowskip=7pt
}

\usepackage[export]{adjustbox}
\usepackage{threeparttable}
\usepackage{multirow}
\usepackage{enumitem}
\usepackage{newfloat}
\usepackage{listings}
\usepackage{colortbl}
\usepackage{amsfonts}
\usepackage{amssymb}
\usepackage{pifont}
\usepackage{float}
\usepackage{bm}

\usepackage{tcolorbox}
\tcbuselibrary{theorems}
\definecolor{lightgray}{gray}{.9}
\definecolor{deepgray}{gray}{.8}
\tcbset{highlight math/.append style={left=0mm,right=0mm,top=0mm,bottom=0mm, colframe=white}}

\usepackage{amsmath}
\usepackage{amssymb}
\usepackage{mathtools}
\usepackage{amsthm}

\usepackage[table]{xcolor} 
\usepackage{amssymb} 

\definecolor{darksalmon}{rgb}{0.98, 0.44, 0.26}
\definecolor{green(pigment)}{rgb}{0.0, 0.65, 0.31}
\definecolor{CustomAvg}{RGB}{149, 159, 105}
\definecolor{CadetBlue}{RGB}{95,158,160} 
\definecolor{DarkBlue}{RGB}{64,101,149}

\newcommand{\greenup}[1]{_{\color{green(pigment)}\uparrow #1}}
\newcommand{\redflat}[1]{_{\color{darksalmon}- #1}}

\title{Self-Spec Verifiable Code Generation}

\author{
Jiaru Qian\textsuperscript{1,4,5,6},
Yihong Dong\textsuperscript{2},
Yongmin Li\textsuperscript{1,4,5},
Hao Zhu\textsuperscript{1,4,5},
Bin Gu\textsuperscript{3,4},
Ge Li\textsuperscript{1}
\\
\textsuperscript{1}School of Computer Science, Peking University
\\
\textsuperscript{2}Shanghai Jiao Tong University
\\
\textsuperscript{3}Beijing Institute of Control Engineering
\\
\textsuperscript{4}Beijing Key Laboratory of Trustworthy Code Large Language Models
\\
\textsuperscript{5}Key Laboratory of High Confidence Software Technologies, Peking University, Ministry of Education
\\
\textsuperscript{6}aiXcoder
\\
}

\iclrfinalcopy 
\begin{document}

\maketitle

\begin{abstract}
Large language models (LLMs) may generate unreliable code on corner cases missed by testing, while formal verification can provide machine-checkable guarantees. 
Recently, researchers have proposed several benchmarks to evaluate the capabilities
of LLMs in generating formally verifiable code, where LLMs need to formulate formal specifications, generate the corresponding code, and verify its correctness.
However, existing benchmarks have two key limitations:
(I) They primarily evaluate specification and code generation stage-wise, with code generation typically conditioned on an oracle specification. This setup overlooks whether strong stage-wise performance translates into end-to-end success.
(II) They mainly focus on a single proof-oriented language and mathematically structured tasks, offering limited coverage of tasks common in software development.
In this paper, we introduce \textsc{VeriCodeBench}, a benchmark for \underline{self-spec} verifiable code generation, where the LLM relies solely on its own generated specification and code throughout the entire process.  \textsc{VeriCodeBench} contains 400 language-native problems across C, Java, Rust, and Python, covering practical concerns in software development. We evaluate specification coverage, code validity, and joint problem-level success across four representative LLMs. We further introduce \textsc{CodeNova} to enhance the capabilities of LLMs in self-spec verifiable code generation. \textsc{CodeNova} makes requirements explicit through constraint-guided specification and uses verifier feedback to guide targeted implementation repairs. 
Experimental results reveal that self-generated specifications remain a major bottleneck, while providing more sophisticated specifications may not necessarily lead to higher verification success rates. \textsc{CodeNova} substantially improves performance across all evaluation metrics, enabling Claude Sonnet 5 to achieve the strongest results under the self-spec protocol.
Our code is available at \url{https://github.com/JiaruQian/VeriCodeBench}.
\end{abstract}



\section{Introduction}
\label{sec: intro}

Large language models (LLMs) are widely used for code generation~\citep{wang2023review,dong2025survey,wang2025planning}. To assess the correctness of LLM-generated code, existing evaluation methods primarily rely on test cases~\citep{ryan2024code,nunez2024autosafecoder}.
However, test cases can cover only a limited portion of the input space and therefore cannot guarantee the correctness of code under all possible circumstances, such as rare inputs, unsafe memory states, or unusual exceptional behavior. Formal verification~\citep{hasan2015formal,wang2025supporting} provides a potential solution by checking programs against formal specifications, thereby offering machine-checkable guarantees for all states that satisfy the specifications.
With the increasing capabilities of LLMs, generating code that is formally verified by construction has emerged as a promising possibility~\citep{aggarwal2024alphaverus,dougherty2025proving}. Starting from user requirements, an LLM needs to formulate formal specifications, generate the corresponding code, and verify its correctness:
\[
\text{requirement}\;\longrightarrow\;\text{specification}\;\longrightarrow\;
\text{code}\;\longrightarrow\;\text{verification}.
\]
We refer to this process as \emph{verifiable code generation}. 
Heretofore, several benchmarks~\citep{thakur2026clever,ye2026verina,le2025formalbench} have been proposed to evaluate the capabilities of LLMs in verifiable code generation. However, existing evaluations leave two complementary gaps. 

First, prevailing benchmarks decouple specification generation from downstream code generation and verification. They then simply combine results from these stages to assess the model's capability for "end-to-end" verifiable code generation. Such stage-wise protocols are useful for diagnosing individual capabilities, but their combination does \textbf{not} indicate the end-to-end capability. Crucially, these stages are \textbf{interdependent}. Errors or design choices in a generated specification can alter the difficulty and even the objective of downstream code generation. Conversely, a more complete specification may impose stronger proof obligations and reduce verification success. Consequently, strong specification generation and oracle-spec code generation do not necessarily imply strong end-to-end performance. Evaluating this interaction requires propagating the model-generated specification through the remainder of the pipeline and measuring the resulting joint success.

Second, existing methods and benchmarks are concentrated in a single formal ecosystem, often using proof-oriented languages and mathematically structured tasks. Retrofitting an existing benchmark can instantiate our self-spec protocol, but doing so would still inherit its task distribution, formal language, and verification ecosystem.
Such benchmarks enable careful study of proof synthesis, but provide limited exposure to the programming abstractions and failure modes encountered in software development (e.g., pointer validity and frame conditions in C, object mutation and exceptions in Java). We need a common evaluation shape that is end-to-end by construction while remaining faithful to each language and verification toolchain.

To address the aforementioned gaps, we introduce \textsc{VeriCodeBench}, a multilingual benchmark centered on self-spec verifiable code generation where the LLM relies solely on its own generated artifacts throughout the entire process. \textsc{VeriCodeBench} comprises 400 language-native problems across four developer-facing languages (C, Java, Rust, and Python) paired with their native verification ecosystems. 
We separately evaluate whether the generated specification covers requirement-level obligations and whether the resulting code verifies against that exact specification. Their conjunction defines problem-level success. The evaluation process is fully automated and deterministic. \textsc{VeriCodeBench} reports specification coverage, code validity, and joint problem-level success.
We additionally evaluate stage-wise settings to quantify the performance gap between oracle-guided and self-spec generation.
Moreover, the multilingual design of \textsc{VeriCodeBench} serves to broaden coverage to developer-facing programming models and practical software concerns that arise in real-world software development. The four language tracks preserve language-specific syntax, semantics, and proof obligations, exposing models to distinct challenges.

\begin{table}[t]
\centering
\caption{Comparison with representative methods and benchmarks. \emph{Joint Generation} indicates
that a method or benchmark covers both specification and code
generation.
\emph{Self-Spec}
means that LLM relies solely on its own generated specification rather than an oracle throughout the entire process.}
\label{tab:benchmark-comparison}
\small
\setlength{\tabcolsep}{3.0pt}
\renewcommand{\arraystretch}{0.85}
\resizebox{\textwidth}{!}{%
\begin{tabular}{lccccc}
\toprule
Method / Benchmark & Joint Generation & Self-Spec & Multilingual & Language & Size \\
\midrule
nl2spec~\citep{cosler2023nl2spec} & $\times$  & $\times$ & $\times$ & LTL & 36 \\
AutoSpec~\citep{wen2024enchanting} & $\times$  & $\times$ & $\times$ & C  & 251 \\
SpecGen~\citep{ma2025specgen} & $\times$  & $\times$ & $\times$ & Java  & 385 \\
ClassInvGen~\citep{sun2025classinvgen} & $\times$  & $\times$ & $\times$ & C++ & 9 \\
PropertyGPT~\citep{liu2024propertygpt} & $\times$  & $\times$ & $\times$ & Solidity & 23 \\
SLD-Spec~\citep{chen2025sld} & $\times$  & $\times$ & $\times$ & C  & 62 \\
WybeCoder~\citep{gloeckle2026wybecoder} & $\times$  & $\times$ & $\times$ & Lean & 360 \\
\addlinespace[2pt]
\midrule
Dafny-Synthesis~\citep{misu2024towards} & $\checkmark$  & $\times$ & $\times$ & Dafny & 153 \\
AlgoVeri~\citep{zhao2026algoveri} & $\times$  & $\times$ & $\checkmark$ & Dafny, Verus, Lean & 77 \\
CLEVER~\citep{thakur2026clever} & $\checkmark$ & $\times$ & $\times$ & Lean & 161 \\
VERINA~\citep{ye2026verina} & $\checkmark$  & $\times$ & $\times$ & Lean & 189 \\
\midrule
\textbf{\textsc{VeriCodeBench} (ours)} & $\checkmark$  & $\checkmark$ & $\checkmark$ & \textbf{C, Java, Rust, Python} & \textbf{400} \\
\bottomrule
\end{tabular}%
}
\vspace{-10pt}
\end{table}


The benchmark also motivates a practical approach to improving verifiable code generation. We introduce \textsc{CodeNova}\footnote{NOVA stands for Natural-language Objectives to Verified Artifacts.}, an end-to-end approach to verifiable code generation from natural-language requirements. \textsc{CodeNova} combines requirement formalization, code generation, and verifier-guided repair in a unified generation pipeline. First, \textsc{CodeNova} extracts atomic behavioral and safety constraints from the requirement, translates them into the target specification language, and self-checks the resulting specification for requirement coverage. The pipeline then generates an implementation conditioned on this specification. Then, \textsc{CodeNova} uses verifier feedback to analyze unsatisfied proof obligations, generate focused repair candidates, and check them with the native verifier. Overall, \textsc{CodeNova} addresses requirement formalization and iterative implementation refinement to improve verifiable code generation under the self-spec protocol.

Our main contributions are three-fold:
\begin{itemize}
  \setlength{\itemsep}{2pt}
  \setlength{\parskip}{0pt}
  \setlength{\parsep}{0pt}
  \item We present \textsc{VeriCodeBench}, a 400-problem benchmark covering programming abstractions and verification challenges common in software development. Its multilingual tracks use native verification toolchains and capture language-specific concerns.
  \item We define self-spec verifiable code generation, preserving
  the generated specification through code generation, repair, and verification. We evaluate the whole pipeline with separate specification-adequacy and code-validity measurements.
  \item We introduce \textsc{CodeNova}, an end-to-end generation approach
  combining constraint-guided specification with verifier-guided candidate
  repair. Experimental results demonstrate the effectiveness of \textsc{CodeNova}.
\end{itemize}

\section{Related Work}
\label{sec:related-work}

\subsection{Benchmarks for Formal Reasoning and Verification}

Benchmarks for formal reasoning address different parts of the path from
informal intent to machine-checked correctness. MiniF2F~\citep{zheng2021minif2f} evaluates theorem
proving on competition-level mathematics across multiple formal systems, while ProofNet~\citep{azerbayev2023proofnet} pairs undergraduate mathematics problems with
informal proofs and Lean statements to support both autoformalization and
formal proving. These benchmarks test mathematical reasoning
but do not center on synthesizing executable programs with state and memory
obligations. 

FormalBench~\citep{le2025formalbench} evaluates specification inference for existing programs, while AlgoVeri~\citep{zhao2026algoveri} aligns classical algorithms across Dafny, Verus, and Lean under equivalent functional contracts. CLEVER~\citep{thakur2026clever} separates specification generation, equivalence checking, implementation synthesis, and correctness proofs on Lean-adapted HumanEval problems. VERINA~\citep{ye2026verina} supports generation of code, specifications, and proofs on curated Lean tasks.
However, these benchmarks primarily evaluate individual stages of the verifiable code generation pipeline rather than the model's ability to autonomously complete the full process. Their application scenarios are also relatively narrow and do not fully reflect the abstractions and failure modes of real-world software development.
\textsc{VeriCodeBench} addresses these gaps with a self-spec setting, where the model relies solely on its own generated specification and code throughout the full verifiable code generation pipeline. Moreover, \textsc{VeriCodeBench} contains 400 language-native tasks across C, Java, Rust, and Python, covering common programming scenarios and language-specific verification challenges.

\subsection{LLM-Assisted Formal Verification}

LLMs have been applied to several tasks in formal verification, including
specification generation, program synthesis, and proof generation.
Work in Dafny, Why3, Lean, and related
systems~\citep{baksys2025minif2fdafny,zheng2021minif2f,azerbayev2023proofnet,yang2023leandojo}
has explored LLM prompting, proof search, and verifier-guided repair
to produce machine-checked programs or proofs.
Recent efforts~\citep{wen2024enchanting,ma2025specgen,le2025formalbench,wu2025specification}
also target developer-facing verification ecosystems, including
ACSL/Frama-C, JML/OpenJML, and Verus, using verifier feedback to
refine specifications, implementations, or proof annotations.
However, most prevailing systems only focus on individual stages of the
verifiable code generation pipeline,
without integrating these stages into a unified workflow.
Our \textsc{CodeNova} combines requirement formalization, code generation, and verifier-guided repair in a unified generation pipeline, improving performance across the full verifiable code generation process.




\section{Self-Spec Verifiable Code Generation}
\label{section: Self-Spec}

\begin{figure}
    \centering
    \includegraphics[width=1.0\textwidth]{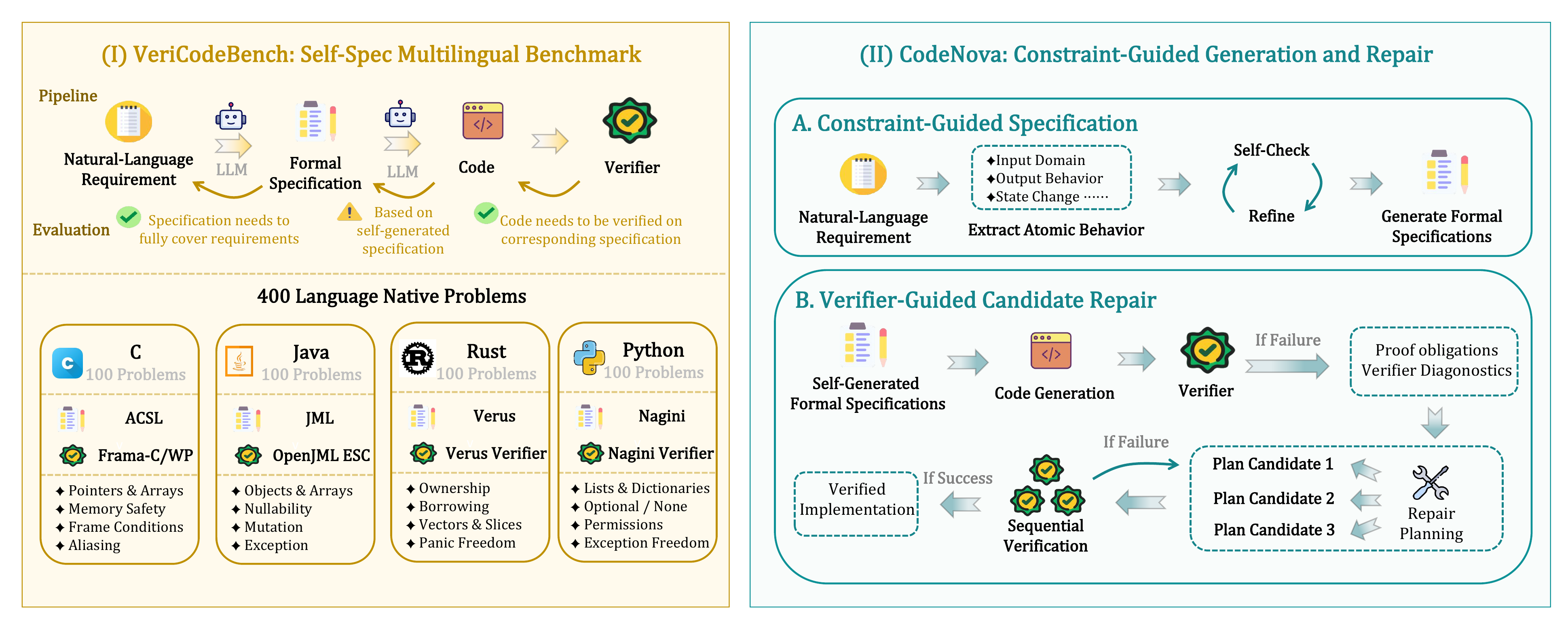}
    \caption{Overview of \textsc{VeriCodeBench} and \textsc{CodeNova}.
    \textbf{Left:} \textsc{VeriCodeBench} evaluates the self-spec pipeline by
    jointly measuring whether the generated specification covers the
    requirement and whether the generated code verifies
    against that same specification, across 400 language-native problems. \textbf{Right:} \textsc{CodeNova} combines
    Constraint-Guided Specification with Verifier-Guided
    Candidate Repair to generate verified code under
a self-generated specification.}
    \label{fig:method-overview}
\end{figure}

\subsection{Problem Formulation}
\label{sec:problem-formulation}

Let $\ell$ denote a programming-language and verification-toolchain track (e.g., C with ACSL and Frama-C/WP). A benchmark instance is a tuple
\begin{equation}
  x_i^{\ell} = (r_i^{\ell}, h_i^{\ell}, G_i^{\ell}, c_i^{\star,\ell},
  V^{\ell}),
\end{equation}
where $r_i^{\ell}$ is a natural-language requirement, $h_i^{\ell}$ is the
provided function signature and language context, $G_i^{\ell}$ is a curated set
of target specification obligations, $c_i^{\star,\ell}$ is a verifier-accepted
reference implementation, and $V^{\ell}$ is the language-native verifier. The
reference specification and implementation define and validate the benchmark
instance; \textbf{neither} is exposed to the model.

Given only $(r_i^{\ell},h_i^{\ell})$, a system first generates a
language-specific specification
\begin{equation}
  \hat{s}_i^{\ell} = F_{\mathrm{spec}}^{\ell}
  (r_i^{\ell},h_i^{\ell}),
  \label{eq:spec-generation}
\end{equation}
and then generates an implementation conditioned on that specification,
\begin{equation}
  \hat{c}_i^{\ell} = F_{\mathrm{code}}^{\ell}
  (r_i^{\ell},h_i^{\ell},\hat{s}_i^{\ell}).
  \label{eq:code-generation}
\end{equation}
The resulting task is therefore not two independent predictions, but the
coupled computation
\begin{equation}
  (r_i^{\ell},h_i^{\ell})
  \longrightarrow \hat{s}_i^{\ell}
  \longrightarrow \hat{c}_i^{\ell}
  \longrightarrow V^{\ell}(\hat{c}_i^{\ell},\hat{s}_i^{\ell}).
  \label{eq:e2e-pipeline}
\end{equation}
This dependency is essential: the specification is both a prediction whose
semantic adequacy must be assessed and an intermediate artifact that constrains
downstream code generation.

\subsection{VeriCodeBench}
\label{VeriCodeBench}

\textsc{VeriCodeBench} operationalizes the task defined above through three
complementary design choices. First, its self-spec evaluation protocol
requires the model to rely exclusively on its own generated specifications
throughout code generation, repair, and verification, without access to
oracle specifications.
Second, it assesses specification adequacy and code validity separately.
Specification adequacy is scored by a fully automated, deterministic
procedure, without LLM-based judging or subjective human assessment.
Joint success requires both full coverage of curated requirement-level
obligations and verifier acceptance under the generated specification for the same
problem. Third, it comprises four language-native tracks with 400
problems, evenly distributed across C, Java, Rust, and Python. The tracks share
this evaluation protocol while retaining distinct programming abstractions,
specification languages, and verification toolchains. Together, these choices enable
end-to-end evaluation of whether models can faithfully formalize requirements
and produce implementations that verify against their own specifications across
diverse programming settings.

\subsubsection{Self-Spec Evaluation}
\label{sec:self-spec-evaluation}
We call a pipeline self-spec when the specification generated in
Equation~\ref{eq:spec-generation} is the mandatory specification supplied in
Equation~\ref{eq:code-generation}. The
pipeline may add implementation-level proof annotations or repair the function
body. 
This is the primary evaluation setting in \textsc{VeriCodeBench}. It reflects
the information boundary of autonomous use: after receiving a natural-language
request, the system cannot assume access to a human-written formalization that
corrects its own interpretation. It also prevents an oracle specification from
masking error propagation. For example, code may verify against a generated
specification that omits an important behavior, while a specification that excludes valid
inputs may make verification artificially easy. Both outcomes are failures of
the end-to-end task even when the verifier accepts the implementation. Moreover,
we also conduct stage-wise diagnostic evaluation as a secondary analysis.
We replace $\hat{s}_i^{\ell}$ in
Equation~\ref{eq:code-generation} with the curated oracle specification
$s_i^{\star,\ell}$. This controlled setting estimates code-generation ability
when specification error is removed and helps attribute the gap between oracle
and self-spec performance.

\subsubsection{Two Independent Correctness Condition}
\label{sec:correctness-conditions}
Verifier acceptance establishes that code satisfies a specification, not that the
specification faithfully captures the requirement. We therefore evaluate two
conditions separately. First, specification adequacy compares the
generated specification $\hat{s}_i^{\ell}$ with the manually curated,
requirement-level target set $G_i^{\ell}=\{g_{i,1}^{\ell},\ldots,g_{i,M_i}^{\ell}\}$.
This comparison uses the Constraint Entailment Framework (CEF), which
combines fixed, language-specific normalization and matching rules with
SMT-based entailment checks. Given the curated targets and a generated
specification, adequacy scoring is fully automated and deterministic: it requires
neither LLM calls nor human adjudication. 

A generated specification may split, merge, or restructure target clauses, so CEF
tests each target against conjunctions of compatible generated clauses. For a
nonempty subset $U$ of normalized clauses, write
$\Phi(U)=\bigwedge_{\hat{s}\in U}\hat{s}$. Coverage uses the entailment
direction below:
\begin{equation}
  \operatorname{Covered}(g_{i,m}^{\ell}) =
  \begin{cases}
    1, & \exists U:\; g_{i,m}^{\ell}\models\Phi(U),
      \quad g_{i,m}^{\ell}\text{ is a precondition},\\
    1, & \exists U:\; \Phi(U)\models g_{i,m}^{\ell},
      \quad g_{i,m}^{\ell}\text{ is a post-, frame-, or exceptional condition},\\
    0, & \text{otherwise}.
  \end{cases}
  \label{eq:specification-entailment}
\end{equation}
The reversed direction for preconditions prevents credit for an unnecessarily
restrictive domain; guarantees must imply the target behavior. CEF returns the
fraction of curated targets with a witness,
\begin{equation}
  A_i^{\ell} = \frac{1}{M_i}
  \sum_{m=1}^{M_i}\operatorname{Covered}(g_{i,m}^{\ell}).
  \label{eq:specification-adequacy}
\end{equation}
The denominator $M_i$ is fixed, so missing, unparsable, or unsupported
specifications receive zero credit for affected targets. Search and entailment
implementation details appear in Appendix~\ref{app:cef-details}.

Second, code validity asks whether the language-native verifier accepts
the generated implementation against the exact generated specification ($\mathbb{1}$ denotes the indicator function):
\begin{equation}
  C_i^{\ell} =
  \mathbb{1}\!\left[V^{\ell}
  (\hat{c}_i^{\ell},\hat{s}_i^{\ell})=\mathrm{valid}\right].
  \label{eq:code-validity}
\end{equation}
Syntax errors, failed proof obligations, verifier errors, and timeouts are not
counted as valid. The primary problem-level measure requires both conditions to
hold for the same benchmark instance:
\begin{equation}
  J_i^{\ell} =
  \mathbb{1}\!\left[
    C_i^{\ell}=1 \;\land\; A_i^{\ell}=1
  \right].
  \label{eq:joint-success}
\end{equation}
Thus, \emph{joint success} requires a verifier-accepted implementation and full
coverage of the target requirement on the same problem. \textsc{VeriCodeBench} reports 
mean specification coverage,
code-validity rate, and joint-success. This metric suite distinguishes
an implementation failure from a specification failure and rejects verification
success obtained under a vacuous specification.

\subsubsection{Four Language Tracks}
\label{sec:language-tracks}
\begin{table}[h]
\centering
\caption{The four language-native tracks in \textsc{VeriCodeBench}. Each track
contains 100 problems.}
\label{tab:language-tracks}
\small
\setlength{\tabcolsep}{4.5pt}
\renewcommand{\arraystretch}{1.0}
\resizebox{\textwidth}{!}{%
\begin{tabular}{llllp{7.1cm}}
\toprule
Language & Specification & Verifier & Size & Native coverage \\
\midrule
C & ACSL & Frama-C/WP  & 100 &
Pointers and aliasing, arrays and slices, loops, byte/string buffers, structs,
frame conditions, and bounded C arithmetic \\
Java & JML & OpenJML ESC & 100 &
Arrays, scalar and Boolean APIs, strings, nullability, object invariants,
object/field frames, mutation, and exceptional behavior \\
Rust & Verus & Verus & 100 &
Scalar arithmetic, \texttt{Option}/\texttt{Result}, vectors and slices,
ownership and borrowing, unique mutation, bounds, and panic freedom \\
Python & Nagini specifications & Nagini & 100 &
Typed scalar functions, \texttt{Optional}/\texttt{None}, lists and dictionaries,
predicate permissions, mutation, and exception freedom \\
\bottomrule
\end{tabular}%
}
\vspace{-8pt}
\end{table}
\noindent \textbf{C/ACSL/Frama-C.}
The C track covers 13 categories, including pointer and pointer-block
manipulation, mutable and immutable arrays, array slices, loops, byte and string
buffers, records, and scalar arithmetic. Specifications use ACSL
\texttt{requires}, \texttt{ensures}, and \texttt{assigns} clauses~\citep{baudin2008acsl} , and
implementations are checked with Frama-C's weakest-precondition plugin . The
track makes memory-side conditions explicit: functional behavior interacts with
pointer validity, separation, buffer bounds, frame conditions, loop invariants,
and machine-integer safety. These obligations make C substantially different
from a purely algorithmic synthesis benchmark.

\noindent \textbf{Java/JML/OpenJML.}
The Java track spans 15 categories centered on arrays, scalar arithmetic,
Boolean logic, strings and characters, mutation, nullability, object state, and
exceptions. JML specifications~\citep{burdy2005overview} are checked by OpenJML~\citep{cok2011openjml} in extended static checking
mode. Problems involving helper classes additionally provide a fixed
\emph{type context} that declares available fields, visibility, and class
invariants. This context prevents the model from inventing a different
object representation without revealing the curated JML specification. The track
therefore tests both functional postconditions and Java-specific behavioral
specifications such as \texttt{assignable}, object invariants, and exceptional
outcomes.

\noindent \textbf{Rust/Verus.}
The Rust track contains scalar-arithmetic, \texttt{Option}/\texttt{Result},
immutable-vector, vector-mutation, and ownership/slice families, together with
extended variants of each family. Verus specifications ~\citep{lattuada2023verus} are attached to Rust
function signatures through \texttt{requires} and \texttt{ensures} clauses.
The problems emphasize safe-Rust functional correctness while retaining
verification challenges not present in ordinary compilation: view-based
reasoning about vectors and slices, pre/post-state relations for mutable borrows,
bounds and panic freedom, and proof annotations for loops. In contrast to the C
track, ownership and borrowing rule out broad classes of aliasing behavior but
introduce their own specification vocabulary and proof obligations.

\noindent \textbf{Python/Nagini.}
The Python track targets the typed subset supported by Nagini ~\citep{eilers2018nagini}. Its 12 categories
cover scalar functions, \texttt{Optional}/\texttt{None},  mutable
list operations, dictionary APIs, and exception freedom, including extended
variants. specifications are expressed as executable-looking
\texttt{Requires(...)} and \texttt{Ensures(...)} statements, while Nagini
translates verification conditions to its underlying permission logic. Container
problems therefore include predicate permissions such as list or dictionary
access, and use pre-state expressions when postconditions refer to values before
mutation. This track tests whether models can produce statically verifiable
specifications and implementations despite Python's otherwise dynamic surface
syntax.

\subsection{CodeNova}
\label{sec:method}
\textsc{CodeNova} combines Constraint-Guided Specification (CGS) with
Verifier-Guided Candidate Repair (VGCR) to generate verified code under
a self-generated specification. Given a requirement $r$ and fixed interface $h$,
CGS produces a specification $\hat{s}$; VGCR then repairs its implementation using
the native verifier $V$. Appendix~\ref{app:codenova-details} provides implementation
details.

\subsubsection{Constraint-Guided Specification}
\label{sec:cgs}

CGS separates requirement interpretation from formalization. It first extracts
atomic behavioral and safety constraints $\mathcal{Q}=\{q_1,\ldots,q_m\}$,
covering admissible inputs, outputs, state changes, and language-specific
obligations, then translates them into a specification:
\begin{equation}
  \mathcal{Q}=F_{\mathrm{extract}}(r,h),
  \qquad s^{(0)}=F_{\mathrm{translate}}(r,h,\mathcal{Q}).
  \label{eq:cgs-translation}
\end{equation}
The model reviews the specification against the requirement and extracted
constraints, identifying omissions and inconsistencies. Its assessment
$u^{(j)}$ guides refinement when needed:
\begin{equation}
  \begin{aligned}
    u^{(j)} &= F_{\mathrm{check}}(r,h,\mathcal{Q},s^{(j)}),\\
    s^{(j+1)} &= F_{\mathrm{refine}}(r,h,\mathcal{Q},s^{(j)},u^{(j)}).
  \end{aligned}
  \label{eq:cgs-refinement}
\end{equation}
Review stops when the model reports alignment or no remaining issues, or the
review budget is exhausted; the selected specification is saved as $\hat{s}$.
This self-check is a generation heuristic: $\mathcal{Q}$ is model-generated, and CEF independently evaluates $\hat{s}$ without external
hints. Auxiliary proof hints, such as loop
invariants, are kept separate from the function specification and remain mutable.

\subsubsection{Verifier-Guided Candidate Repair}
\label{sec:vgcr}

The pipeline generates $c_0=F_{\mathrm{code}}(r,h,\hat{s})$ and checks
$V(c_0,\hat{s})$. On failure, VGCR uses verifier diagnostics $d_t$ for the
current implementation $c_t$ to plan repairs and propose up to $K$ candidates:
\begin{align}
  P_t &= F_{\mathrm{plan}}(r,h,\hat{s},c_t,d_t),
  \label{eq:vgcr-plan}\\
  \tilde{c}_{t,k} &= F_{\mathrm{repair}}
    (r,h,\hat{s},c_t,P_t,f_{t,k}), \quad k=1,\ldots,K.
  \label{eq:vgcr-candidate}
\end{align}
The plan $P_t$ decomposes failures into subgoals with diagnostic evidence and
repair hints, such as correcting bounds checks or strengthening loop
invariants. The focus $f_{t,k}$ directs a candidate toward combined subgoals,
an individual issue, or an alternative implementation. Candidates share the
same starting code and plan within a round, and are checked sequentially as
complete programs by $V$. 
VGCR returns the first verifier-accepted candidate. If none passes, the last
candidate submitted to $V$ and its diagnostics seed the next round; candidates
rejected by language-subset prechecks do not replace the current code.
If the repair budget is exhausted without verifier acceptance, the
instance is recorded as a failure. By translating verifier diagnostics
into explicit repair subgoals, VGCR guides changes to executable code
and local proof annotations. Exploring candidates with different repair
focuses provides alternative ways to resolve verification failures,
while checking each candidate as a complete program validates whether
the proposed changes satisfy the proof obligations.

\section{Experiments}
\label{sec:experiments}

In this section, we conduct thorough evaluations of frontier LLMs and \textsc{CodeNova} on \textsc{VeriCodeBench}. We organize the evaluation around four research questions.
\textbf{RQ1 (Section \ref{sec:main-results}):} How well do current LLMs and \textsc{CodeNova} perform on end-to-end verified 
code generation under the self-spec protocol? 
\textbf{RQ2 (Section \ref{sec:oracle-diagnostic}):} How does the source of the specification affect downstream code generation?
\textbf{RQ3 (Appendix~\ref{app:passk-analysis}):} How does joint success scale with the sampling budget, as measured by pass@$k$?
\textbf{RQ4 (Appendix ~\ref{sec:case-study}):} What do representative case studies reveal about the strengths and limitations in end-to-end verified code generation?

\subsection{Experimental Setup}
\label{sec:experimental-setup}

\noindent \textbf{Models and Inference Configuration.}
We evaluate DeepSeek V3.2, Kimi-K2.7-Code, Qwen3.6-plus, and Claude-Sonnet-5
on all 400 problems. Each pipeline uses the same base LLM for specification
generation, code generation, and any subsequent review or repair. We also evaluate the performance with different sampling budgets on DeepSeek V4.1 Flash.

\noindent \textbf{Implementation Details.}
We verify each implementation using the native verification toolchain for its respective language: Frama-C/WP for C,
OpenJML for Java, Verus for Rust, and Nagini for Python. Code validity requires
verifier acceptance under the specification supplied to generation. Syntax or type
errors, unproved obligations, verifier errors, and timeouts count as failures.
The primary protocol is self-spec.
We compare four configurations: \textbf{Direct} generates a specification and
one implementation without repair; \textbf{CGS} replaces direct specification
generation with Constraint-Guided Specification; \textbf{VGCR} adds
Verifier-Guided Candidate Repair to Direct; and \textbf{\textsc{CodeNova}}
combines CGS and VGCR. Implementation details are described in
Appendix~\ref{sec:codenova-ablation}.
We report mean per-problem requirement coverage (Req.\ cov.), the number of
verifier-accepted implementations (Valid), and the number that additionally
achieve complete requirement coverage (Joint), following
Section~\ref{sec:problem-formulation}. Each track contains 100 problems, so
Valid and Joint counts also equal percentage rates.

\subsection{End-to-End Self-Spec Performance}
\label{sec:main-results}

\begin{table}[t!]
\centering
\caption{Results on \textsc{VeriCodeBench}, with 100 problems
per language. Joint denotes joint-success counts;
Req.\ cov. represents mean requirement coverage; Valid means
code-valid counts. Direct and VGCR share the same generated
specifications, as do CGS and \textsc{CodeNova}. Joint success requires both
complete requirement coverage and verifier acceptance on the same problem. \textbf{Bold} values denote the best result. Second best results are \underline{underlined}.}
\label{tab:main-results}
\small
\setlength{\tabcolsep}{3pt}
\resizebox{\textwidth}{!}{%
\begin{tabular}{ll|ccc|ccc|ccc|ccc}
\toprule
\multirow{2}{*}{Base LLM} & \multirow{2}{*}{Method} & \multicolumn{3}{c}{C (100)} & \multicolumn{3}{c}{Java (100)} & \multicolumn{3}{c}{Rust (100)} & \multicolumn{3}{c}{Python (100)} \\
 & & Joint & Req. cov. & Valid & Joint &  Req. cov. & Valid & Joint & Req. cov. & Valid & Joint & Req. cov. & Valid  \\
\midrule
\multirow{4}{*}{DeepSeek V3.2}
 & Direct & 8 & 0.3247 & 34 & 30 & 0.7113 & 76 & 33 & 0.5522 & 44 & 46 & 0.8267 & 68  \\
 & +VGCR & 11 & 0.3247 & 60 & 31 & 0.7113 & 82 & 45 & 0.5522 & 68 & 46 & 0.8267 & 71  \\
  & +CGS & 12 & 0.5215 & 30 & 36 & 0.7810 & 56 & 30 & 0.5572 & 49 & 51 & 0.8593 & 65  \\
 & \textsc{CodeNova} & 17 & 0.5215 & 54 & 38 & 0.7810 & 66 & 35 & 0.5572 & 58 & 52  & 0.8593 & 68 \\
\midrule
\multirow{4}{*}{Kimi-K2.7-Code}
 & Direct& 16 & 0.5856 & 68 & 56 & 0.8250 & 77 & 57 & 0.6965 & 91 & \underline{76} & 0.9327 & \underline{86}  \\
 & +VGCR & 17 & 0.5856 & \textbf{86} & 59  & 0.8250 & 86 & 57 & 0.6965 & 92 & \underline{76}  & 0.9327 & \underline{86} \\
  & +CGS & 21 & 0.6529 & 71 & 55 & 0.8265 & 68 & 43 & 0.7562 & 60 & 73  & 0.9380 & 82  \\
 & \textsc{CodeNova} & 22 & 0.6529 & \textbf{86} & 61 & 0.8265 & 79 & 43 & 0.7562 & 60 & 73  & 0.9380 & 82  \\
\midrule
\multirow{4}{*}{Qwen3.6-plus}
 & Direct & 13 & 0.4480 & 50 & 37 & 0.6522 & 83 & 38  & 0.6667 & 56 & 54 & 0.8940 & 65  \\
 & +VGCR & 17 & 0.4480 & 64 & 37 & 0.6522 & 91  & 50 & 0.6667 & 76 & 55  & 0.8940 & 66 \\
  & +CGS & 18 & 0.6552 & 39 & 46 & 0.7958 & 82  & 42 & 0.7164 & 54 & 61 & 0.9148 & 71  \\
 & \textsc{CodeNova} & 24 & 0.6552 & 66 & 46 & 0.7958 & 88 & 49 & 0.7164 & 69  & 64 & 0.9148 & 74  \\
\midrule
\multirow{4}{*}{Claude-Sonnet-5}
 & Direct & 30 & \underline{0.6878} & 73 & 65 & \underline{0.9060} & 90 & 68 & \underline{0.8060} & 96 & \underline{76}   & \underline{0.9430} & 79 \\
 & +VGCR & \textbf{33} & \underline{0.6878} & \underline{80} & \underline{71} & \underline{0.9060} & \textbf{99} & 68 & \underline{0.8060} & \textbf{98} & \underline{76}  & \underline{0.9430} & 79  \\
  & +CGS & 23 & \textbf{0.7157} & 56 & 63 & \textbf{0.9097} & 82 & \underline{71}  & \textbf{0.8507} & 94 & \textbf{85} & \textbf{0.9673} & \textbf{90}  \\
 & \textsc{CodeNova} & \underline{31} & \textbf{0.7157} & 77 & \textbf{73}  & \textbf{0.9097} & \underline{97} & \textbf{73} & \textbf{0.8507} & \underline{97} & \textbf{85} & \textbf{0.9673} & \textbf{90}  \\
\bottomrule
\end{tabular}%
}
\vspace{-10pt}
\end{table}

We evaluate four LLMs under the four configurations in
Section~\ref{sec:experimental-setup} on all 400 problems using the
self-spec protocol. Results are shown in Table~\ref{tab:main-results}. 
Under direct setting, code validity averages 71.0\% across the 16
model--language settings, but joint success reaches only 44.1\%.
Under Direct setting,
Claude-Sonnet-5 performs the best, achievint 60.3\% joint success in average
across the four tracks. Thus, producing implementations that both verify and
fully cover the requirements remains challenging even for the strongest
evaluated model.

\noindent \textbf{CGS improves coverage but may increase downstream difficulty.}
CGS improves requirement coverage in all 16 settings, raising the mean from
0.7162 to 0.7761 and indicating more complete formalization of the target
obligations. Yet code validity declines in several settings.
For Kimi-K2.7-Code on Rust, coverage rises from 0.6965 to 0.7562 while valid
implementations drop from 91 to 60. This pattern is consistent with more
demanding specifications making implementation and proof generation harder.
Unnecessary complexity or overly restrictive conditions may further compound
the difficulty. Coverage alone does not establish specification complexity or
rule out over-constraint. Despite this trade-off, CGS improves aggregate joint
success from 44.1\% to 45.6\% in average.

\noindent \textbf{VGCR improves verification under both direct and CGS settings.}
VGCR uses verifier feedback to repair code and implementation-level proof
annotations, including loop invariants. Added to Direct, it raises code
validity and joint success significantly.
VGCR also helps implement CGS-generated specifications: combining two stages
raises validity from 65.6\% to 75.7\% and joint success from 45.6\% to
49.1\%. Thus, \textsc{CodeNova} achieves the best aggregate joint success
among the four configurations, with Claude-Sonnet-5 reaching 65.5\% across
tracks. These gains support the complementarity of specification guidance
and verifier-guided repair.

\noindent \textbf{Specification adequacy limits the end-to-end pipeline.}
Downstream improvements remain bounded by the frozen specification.
For DeepSeek V3.2 on C, VGCR increases valid implementations from 34 to 60,
but joint successes rise from 8 to only 11. Because generation, verification,
and repair cannot restore requirement coverage
when the specification omits an obligation. Even verifier-accepted code 
cannot achieve joint success without an adequate initial specification.

\subsection{Error Propagation and Oracle-Spec Diagnostic}
\label{sec:oracle-diagnostic}

\begin{table}[!t]
\centering
\caption{Code validity with self-generated versus oracle specifications.
\textit{Self} uses the Direct or CGS specification from Table~\ref{tab:main-results};
\textit{Oracle} supplies a curated function specification. Both settings require the model
to generate code and implementation-level proof annotations.}
\label{tab:oracle-results}
\small
\setlength{\tabcolsep}{5pt}
\resizebox{0.8\textwidth}{!}{%
\begin{tabular}{llcccccccc}
\toprule
\multirow{2}{*}{Base LLM} & \multirow{2}{*}{Method}
 & \multicolumn{2}{c}{C (100)} & \multicolumn{2}{c}{Java (100)}
 & \multicolumn{2}{c}{Rust (100)} & \multicolumn{2}{c}{Python (100)} \\
\cmidrule(lr){3-4}\cmidrule(lr){5-6}\cmidrule(lr){7-8}\cmidrule(lr){9-10}
 & & Self & Oracle & Self & Oracle & Self & Oracle & Self & Oracle \\
\midrule
\multirow{2}{*}{DeepSeek V3.2}
 & Direct & 34 & $69\greenup{25}$ & 76 & $94\greenup{18}$ & 44 & $70\greenup{26}$ & 68 & $90\greenup{22}$ \\
 & \textsc{CodeNova}   & 54 & $69\greenup{15}$ & 66 & $99\greenup{33}$ & 58 & $91\greenup{33}$ & 68 & $93\greenup{25}$ \\
\midrule
\multirow{2}{*}{Kimi-K2.7-Code}
 & Direct & 68 & $70\greenup{2}$ & 77 & $96\greenup{19}$ & 91 & $91\redflat{}$ & 86 & $92\greenup{6}$ \\
 & \textsc{CodeNova}   & 86 & $89\greenup{3}$ & 79 & $100\greenup{21}$ & 60 & $93\greenup{33}$ & 82 & $95\greenup{13}$ \\
\midrule
\multirow{2}{*}{Qwen3.6-plus}
 & Direct & 50 & $86\greenup{36}$ & 83 & $93\greenup{10}$ & 56 & $74\greenup{18}$ & 65 & $90\greenup{25}$ \\
 & \textsc{CodeNova}   & 66 & $92\greenup{26}$ & 88 & $99\greenup{11}$ & 69 & $80\greenup{11}$ & 74 & $92\greenup{18}$ \\
\midrule
\multirow{2}{*}{Claude-Sonnet-5}
 & Direct & 73 & $87\greenup{14}$ & 90 & $96\greenup{6}$ & 96 & $98\greenup{2}$ & 79 & $95\greenup{16}$ \\
 & \textsc{CodeNova}   & 77 & $91\greenup{14}$ & 97 & $100\greenup{3}$ & 97 & $100\greenup{3}$ & 90 & $100\greenup{10}$ \\
\bottomrule
\end{tabular}%
}
\vspace{-10pt}
\end{table}

\begin{figure*}[t]
    \centering
    \includegraphics[width=\textwidth]{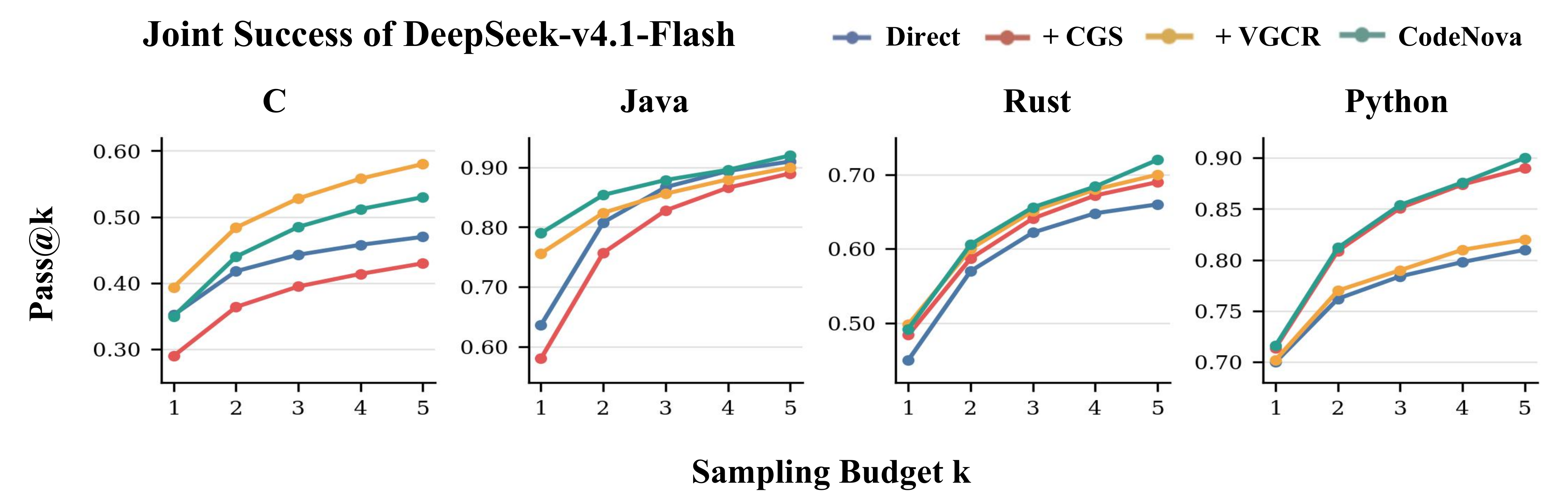}
    \vspace{-20pt}
    \caption{Pass@$k$ joint-success curves across the four language tracks.
    The panels use language-specific y-axis ranges to improve visual resolution.
    Detailed analysis are provided in Appendix~\ref{app:passk-analysis}.}
    \label{fig:pass-at-k}
    \vspace{-10pt}
\end{figure*}

We compare code generation conditioned on the model's own specification
(\emph{Self}) with code generation conditioned on a curated oracle function
specification (\emph{Oracle}). The oracle supplies interface-level obligations,
including applicable preconditions, postconditions, and frame conditions;
the model must still generate the implementation and its implementation-level
proof annotations, such as loop invariants. We evaluate Direct and \textsc{CodeNova} in both
settings. The reported metric
is code validity under the respective supplied specification, not self-spec
joint success.

\noindent \textbf{Oracle specifications reveal substantial hidden difficulty in self-spec generation.}
Table~\ref{tab:oracle-results} shows that oracle specifications improve code validity in nearly every comparison. With oracle specifications, strong models such as Claude-Sonnet-5 achieve consistently high verification success, indicating that downstream code generation and proof synthesis are considerably more capable when the specification is reliable. This explains why existing benchmarks that evaluate oracle-spec code generation separately and combine it with specification generation performance may provide overly optimistic estimates of end-to-end verified code generation.

\noindent \textbf{Code generation performance is sensitive to which specification it receives.}
Even with VGCR, \textsc{CodeNova} using self-generated specifications remains 12.4 percentage points lower than the oracle setting in aggregate validity. This gap demonstrates that improving downstream repair alone cannot fully eliminate the challenges introduced by self-spec generation. However, the gap should not be interpreted as a direct measurement of specification errors, because replacing a self-generated specification with an oracle specification changes both the information available to the model and the verification obligations. Conversely, a small oracle gap does not necessarily imply adequate specifications. For example, Kimi-K2.7-Code on Rust achieves identical Direct validity under Self and Oracle specifications, while its self-spec joint success remains substantially lower. These results highlight that specification quality and verifier acceptance capture different failure modes, and both are necessary for reliable end-to-end verified code generation. Cases can be found in Appendix~\ref{subsec:sensitive}.

\section{Conclusion}
\label{sec:conclusion}

We introduced \textsc{VeriCodeBench}, a self-spec multilingual benchmark of 400 problems across C,
Java, Rust, and Python. By jointly assessing requirement
coverage and native verifier acceptance, the benchmark exposes failures that
verification under a generated specification alone can miss. Our evaluation shows
that specification quality strongly affects downstream verification and
remains a bottleneck for end-to-end success. \textsc{CodeNova} combines
constraint-guided specification with verifier-guided candidate repair to
improve aggregate joint success, while its remaining failures highlight the
limits of repair under an inadequate specification. These findings motivate
methods that advance faithful requirement formalization alongside
implementation and proof generation, and evaluation that preserves their
dependencies throughout the pipeline.

\bibliography{iclr2027_conference}
\bibliographystyle{iclr2027_conference}

\appendix

\section{Implementation Details of VeriCodeBench}

\subsection{Dataset Construction and Quality Control}

We construct every track in four stages. First, each problem is assigned a
concise natural-language requirement and a fixed language-level signature. The
signature removes irrelevant interface search while leaving behavioral
formalization to the model. When a task depends on auxiliary declarations, such
as a Java record-like class, the input also includes the minimal type context
needed to compile and verify the target method. The primary protocol exposes
only the requirement, signature, and such public context; target clauses and
reference code remain hidden.

Second, we manually curate a set of atomic target obligations from the intended
behavior. Each target record stores a stable identifier, clause type, formal
expression, natural-language gloss, and provenance. Depending on the track,
clause types include preconditions, postconditions, frame or assignability
conditions, and exceptional behavior. The target set is not obtained by blindly
extracting every annotation from a reference file. Reference implementations
often contain proof-oriented conditions---for example pointer-validity facts,
permission predicates, overflow guards, or helper invariants---that may be
necessary for a particular proof but are not always part of the requirement's
semantic target. Curators include such conditions only when they are part of the
benchmark's intended specification. Conversely, when a concise requirement leaves a
verification-relevant detail implicit, the reference specification is used to
resolve the instance's operational meaning. The released target clauses make
these decisions explicit and auditable.

Third, every instance is paired with a manually constructed implementation and
full native specification. Reference programs may contain loop invariants, frame
annotations, ownership/view facts, or permission assertions required by their
respective verifier. They are checked with the same verifier family used to
score generated artifacts. This serves two purposes: it establishes that the
task is realizable under the fixed interface, and it validates the interaction
between the curated behavioral targets and the language-specific verification
environment. The reference implementation is never used as a candidate answer
in model evaluation.

Finally, automated dataset checks validate identifier alignment, paths,
signatures, clause schemas, and target counts between the requirement and
ground-truth manifests. The current release contains exactly 100 requirements
and 100 corresponding target records in each track. Reference suites are run in
toolchain-specific containers to control verifier and solver dependencies. We
retain older subsets separately for reproducibility, but all results in this
paper use the 100-problem manifests.

\subsection{Design Principles}
The benchmark is organized around three principles.

\paragraph{Dependency fidelity.}
The benchmark preserves the causal dependency of autonomous generation. A model
first produces a specification and must then implement the program under that same
specification. The specification is treated as a frozen intermediate artifact during code
generation and verifier-guided repair. This prevents a system from obtaining an
easier proof by silently deleting a postcondition, strengthening a precondition,
or substituting the benchmark's target specification. Consequently, the primary
result captures error propagation from requirement interpretation to
specification and finally to verified implementation.

\paragraph{Language-native verification.}
Single-ecosystem benchmarks provide focused measurements within one formal
setting; our multilingual design adds complementary breadth across programming
models and verifier semantics. Rather than translating one abstract task set
into several syntaxes, we evaluate the abstractions developers actually
encounter in each ecosystem. The four tracks share the same high-level
interface---requirement, specification, implementation, verifier, and coverage
targets---but use language-specific problems and native specification idioms. This
choice exposes difficulties such as pointer validity in C, object frames in
Java, ownership-aware mutation in Rust, and permission-based container
reasoning in Python. Cross-language differences should therefore be interpreted
as differences between complete task--toolchain combinations, not as controlled
measurements on parallel translations.

\paragraph{Joint evaluation with diagnostic controls.}
Verifier acceptance and requirement alignment are recorded independently, then
combined only at the problem level as defined in
Equation~\ref{eq:joint-success}. The self-spec path is the benchmark's
primary protocol. At the same time, the curated targets and reference specifications
support oracle-spec and stage-wise controls that localize failures without
changing the primary task. All variants use the same fixed target denominator,
so a generation failure or unparsable specification cannot improve coverage by
removing difficult obligations from evaluation.

\subsection{Constraint Entailment Details}
\label{app:cef-details}

CEF searches compatible generated clauses in increasing conjunction size,
from single clauses through all nonempty subsets, and stops at the first
witness. Thus one target requires at most $2^N-1$ entailment checks. In our implementation, we set $N=5$ and limits the maximum search attempts to $2,000$. Each query
$L\models R$ is reduced to unsatisfiability of $L\land\neg R$ in an SMT solver
after conservative, language-specific canonicalization. Canonicalization
removes superficial differences such as parameter names and supported Boolean
rewrites while preserving pre/post-state distinctions. Exceptional constructs
outside the shared Boolean fragment use conservative type-aware matching. The
LLM is never asked to prove equivalence; coverage depends on the specification logic
itself. Missing, unparsable, or unsupported specifications receive zero credit
for affected targets.

\subsection{Benchmark Artifacts}

\textsc{VeriCodeBench} releases both static dataset artifacts and the machinery
needed to reproduce an end-to-end run. For each problem, the static release
contains the requirement and category metadata, fixed signature and optional
type context, atomic ground-truth obligations, a complete reference specification,
and an annotated reference implementation. Track-level scripts verify reference
programs and invoke the appropriate requirement-to-code pipeline.

For every model run, the pipeline stores four layers of evidence: (i) the
canonical generated specification, including structured clauses and the exact
specification text; (ii) the generated source file; (iii) the normalized verifier
verdict and raw diagnostics; and (iv) the CEF and problem-level summary reports.
The canonical representation is language aware: ACSL and JML use source-level
specification blocks, Verus stores structured signature clauses, and Nagini stores
both specification statements and normalized clauses. This representation is the
source of truth shared by code generation, verification, and coverage scoring.

The pipeline operationalizes the self-spec boundary at the artifact
level. The generated specification is supplied to code generation and remains frozen
during repair; only method bodies, helper proof code, and implementation-level
annotations such as loop invariants may change. The verifier-facing form is
language specific: ACSL and JML blocks appear above the target function, Verus
clauses are rebuilt into the function signature, and Nagini statements appear
at the beginning of the function body. The Java and Python adapters reinsert the
canonical specification after generation and repair, while the Rust adapter derives
the signature from its stored structured clauses; the C pipeline passes the
same stored ACSL block to every downstream prompt. Specification-placement and
enforcement actions are retained in run metadata, and an explicit consistency
check rejects Java source/specification mismatches. Together with raw
diagnostics and fixed manifests, these artifacts make each reported failure
traceable to specification generation, code generation, specification preservation,
or the native verifier.

\section{Implementation Details of CodeNova}
\label{app:codenova-details}

This section supplements the CGS and VGCR procedures in
Section~\ref{sec:method}, including their language-specific representations,
candidate scheduling, and controlled evaluation.

\subsection{Constraint Representation and Specification Refinement}
\label{app:cgs-details}

\paragraph{Atomic constraints.}
Each extracted constraint expresses a single condition in concise mathematical
or logical language. The structured representation groups constraints by role,
including admissible inputs, return-value behavior, state changes, and
invariants. Track-specific fields capture frame conditions and exceptional
behavior in Java, mutation and panic freedom in Rust, and permissions and
exception freedom in Python. The fixed signature and public context constrain
extraction to the supplied interface and object representation. For example,
updating one array element while preserving the others requires both an
updated-value relation and an unchanged-region condition. Distinct return
cases and boundary inputs should likewise remain explicit. The extracted set
$\mathcal{Q}$ can nevertheless be incomplete; it is not the curated target
set $G$ used by CEF.

\paragraph{Native translation and proof hints.}
Translation retains access to the original requirement and interface,
distinguishes input assumptions from output guarantees, and preserves
pre/post-state references for mutation. Language-specific instructions guide
ACSL pointer validity and frames, JML assignability, Verus sequence views and
mutable borrows, and Nagini container permissions. The specification artifact
may also contain auxiliary hints for code generation, including candidate loop
invariants, loop frames, and termination measures where supported. These are
stored separately from the function specification because they depend on a future
implementation and may change during repair. They are neither assumptions on
every function call nor additional requirement-level evaluation targets.

\paragraph{Review and fallback.}
The self-check returns a structured alignment judgment, missing constraints,
inconsistencies, and refinement hints. As in
Equation~\ref{eq:cgs-refinement}, refinement uses those findings to revise the
specification. Review ends when the model reports alignment or lists no
missing or inconsistent items, or when its configured budget is reached.
The adapters validate the specification representation and apply their
supported-syntax checks and fallback rules before saving $\hat{s}$.
Budget exhaustion does not establish alignment. Neither CEF coverage nor its
entailment witnesses are available to this process.

\subsection{Repair Planning and Candidate Scheduling}
\label{app:vgcr-details}

\paragraph{Plan structure.}
The verifier-derived plan contains a failure summary, subgoals, an overall
repair strategy, and potential semantic risks. Each subgoal records an issue
type, supporting diagnostic evidence, and a proposed implementation or
annotation change. Issues can concern syntax and specification placement,
postconditions, memory or permission safety, bounds and overflow, loop
invariant establishment or preservation, frames, or termination. Timeouts and
unclassified failures can also trigger planning. These subgoals are LLM
interpretations of diagnostics, not independently proved lemmas; a failed or
timed-out proof attempt is not treated as a concrete counterexample.

\paragraph{Focus schedule and selection.}
For each plan, the focus schedule first integrates all subgoal hints, then
prioritizes individual subgoals in order, and finally requests an alternative
implementation if the candidate budget reaches that focus. The schedule
repeats if necessary. Each candidate is generated from the same $c_t$ and
$P_t$ within a round and undergoes language-specific processing before native
verification. Checking the complete program preserves interactions between
subgoals: a postcondition repair must still satisfy memory safety, framing,
and all other verifier obligations. The first accepted candidate is returned;
otherwise, the last candidate submitted to the verifier becomes $c_{t+1}$.
Precheck-rejected candidates do not replace the current implementation; if
all candidates are rejected, the round records a failure. Selection uses
verifier acceptance, without an LLM score or an assumption that unproved
obligations decrease monotonically. On budget exhaustion, the final
implementation is retained for analysis.


\subsection{Controlled Component Evaluation}
\label{sec:codenova-ablation}

Direct generates a specification and one implementation, then verifies it
once. CGS changes specification generation while retaining that single code
attempt; VGCR adds repair under Direct's saved specification; full
\textsc{CodeNova} combines both stages. Each repair run reuses the exact
specification and initial code from its corresponding run without repair.
Neither artifact is resampled, so specification coverage is identical within
each pair. This isolates repair gains from specification-generation effects.
Extracted constraints, self-check findings, repair plans, candidate focuses,
and verifier outcomes are recorded for analysis.

\section{\texorpdfstring{Pass@$k$ Analysis}{Pass-at-k Analysis}}
\label{app:passk-analysis}

To address \textbf{RQ3}, we conduct evaluation on the performance with different sampling budgets. We adopt DeepSeek v4.1 Flash as the base model, measuring pass@$k$ performance under four configurations. 
Figure~\ref{fig:pass-at-k} highlights a central distinction between improving
the specification and improving the implementation. CGS can lower the curve
in some language tracks, especially at small sampling budgets, because a more
sophisticated specification may also introduce additional proof obligations or a less
usable proof interface. In other tracks and at larger budgets, however, CGS
raises the attainable success level by making the intended behavior easier to
recover. Its effect is therefore conditional rather than uniformly positive
or negative.

VGCR is more consistently beneficial across the curves. Verifier feedback
provides a mechanism for converting additional samples into targeted repairs,
so the VGCR curves generally dominate their Direct counterparts or approach
the same ceiling more quickly. This effect is complementary to CGS: when CGS
supplies a better-aligned specification, VGCR can spend its repair budget on
implementation and proof obligations rather than compensating for missing
requirements. The combined \textsc{CodeNova} curves consequently recover cases
where CGS alone is harmful while retaining its gains where specification
guidance is useful.

The curves also show why pass@$1$ can understate the potential of the full
pipeline. Several configurations continue to improve as the budget grows,
with \textsc{CodeNova} often retaining useful headroom through pass@$5$.
This pattern is consistent with complementary candidate diversity: different
samples may discover distinct specifications, implementations, or verifier-proof
annotations. At the same time, the flattening of some curves indicates that
sampling cannot remove a persistent specification bottleneck or an
unrealizable obligation. Language-specific verifier semantics and specification
idioms further shape both the initial success rate and the rate of saturation,
so pass@$k$ should be read as a budget-sensitivity diagnostic rather than a
single ranking independent of the track.

\section{Case Study}
\label{sec:case-study}

This section gives artifact-level examples of how constraint extraction and
verification-guided repair affect the two components of joint success. 
Examples are selected from the Claude-Sonnet-5 runs.

\subsection{CGS can repair missing behavioral coverage}

Here CGS makes the specification more explicit while preserving verifier validity,
so a previously incomplete specification becomes a joint success.

\paragraph{C, problem 13 (separate equal and unequal behaviors).}
The base specification expresses equality with one biconditional and covers
only $4/6$ target clauses.  CGS splits the two behaviors into disjoint,
complete ACSL behaviors:

\begin{lstlisting}[style=casecode,language=C]
behavior all_equal:
  assumes \forall integer i; 0 <= i < n ==> a[i] == b[i];
  ensures \result == 1;
behavior not_equal:
  assumes \exists integer i; 0 <= i < n && a[i] != b[i];
  ensures \result == 0;
complete behaviors;
disjoint behaviors;
\end{lstlisting}

The generated code remains valid and coverage rises to $6/6$, producing a
joint success.

\paragraph{Rust, problem 36 (vector mutation).}
The base specification uses one subrange equation and covers $2/3$ targets.  CGS
spells out both the length change and the unchanged prefix:

\begin{lstlisting}[style=casecode]
ensures v.len() == old(v).len() - 1
ensures forall|i: int| 0 <= i < v.len() ==> v[i] == old(v)[i]
\end{lstlisting}

The decomposition matches the evaluator's independent length, changed-element,
and frame targets, giving $3/3$ coverage while retaining verifier acceptance.

\paragraph{Python, problem 57 (complementary branch).}
The base specification covers only the true branch ($1/2$):

\begin{lstlisting}[style=casecode,language=Python]
Ensures(Implies(flag, Result() == x))
\end{lstlisting}

CGS adds the missing false branch, raising coverage to $2/2$:

\begin{lstlisting}[style=casecode,language=Python]
Ensures(Implies(flag, Result() == x))
Ensures(Implies(not flag, Result() == y))
\end{lstlisting}

\subsection{CGS may also make an already-valid program harder to prove}

These cases have full target coverage in both runs, but the richer specification
introduces a proof obligation that the native verifier cannot discharge.

\paragraph{C, problem 5 (pointer aliasing and frames).}
The base program verifies with coverage $2/2$.  CGS adds the semantically useful
frame fact that the read pointer is unchanged:

\begin{lstlisting}[style=casecode,language=C]
/*@ ensures *a == \old(*a) + \old(*b);
    ensures *b == \old(*b); */
\end{lstlisting}

The implementation changes $a$ but says nothing that excludes $a$ and $b$
from aliasing.  Frama-C/WP proves four of five goals and times out on the new
frame goal (10 seconds); the base run proves four of four.  Thus CGS preserves
coverage but changes \texttt{code\_valid} from true to false.

\paragraph{Java, problem 16 (existential loop invariant).}
The base specification and implementation verify with coverage $5/5$.  CGS uses an
existential witness in the loop invariant:

\begin{lstlisting}[style=casecode,language=Java]
/*@ loop_invariant (\exists int k; 0 <= k && k < i && a[k] == min) || i == 0;
  @ loop_invariant \forall int k; 0 <= k && k < i; min <= a[k];
  @*/
\end{lstlisting}

OpenJML cannot establish preservation of the first invariant for the generated
loop.  The specification remains fully covered ($5/5$), but validity is lost because
the proof interface is stronger than the implementation can establish.

\paragraph{Python, problem 94 (length-preservation disjunction).}
The base specification verifies with coverage $4/4$.  CGS adds a length case split:

\begin{lstlisting}[style=casecode,language=Python]
Ensures(key in d)
Ensures(len(d) == Old(len(d)) or
        (key not in Old(d) and len(d) == Old(len(d)) + 1))
Ensures(d[key] == key)
\end{lstlisting}

Nagini fails to prove the disjunctive postcondition for the generated update,
although all four target clauses are covered.  This is a semantic proof
difficulty rather than a parser or type failure.

\subsection{VGCR compensates harder specifications}

The following cases satisfy the strictest preservation test: base and
CGS+VGCR are joint successes, while CGS alone is not.  VGCR edits executable
code or implementation-level annotations only; the CGS specification and its
coverage denominator remain fixed.

\paragraph{Java, problem 16 (explicit witness).}
VGCR replaces the unprovable existential reasoning with a concrete index while
retaining the same $5/5$ specification coverage:

\begin{lstlisting}[style=casecode,language=Java]
int min = a[0];
int minIdx = 0;
int i = 1;
/*@ loop_invariant 0 <= minIdx && minIdx < i;
  @ loop_invariant a[minIdx] == min;
  @ loop_invariant \forall int k; 0 <= k && k < i; min <= a[k];
  @*/
while (i < a.length) {
    if (a[i] < min) { min = a[i]; minIdx = i; }
    i++;
}
\end{lstlisting}

\paragraph{Java, problem 3 (exception guards).}
CGS adds null and empty-array exceptional behaviors.  VGCR makes those cases
explicit before the indexed read, restoring validity and $4/4$ joint success:

\begin{lstlisting}[style=casecode,language=Java]
if (a == null) {
    throw new NullPointerException();
}
if (a.length == 0) {
    throw new ArrayIndexOutOfBoundsException();
}
return a[0];
\end{lstlisting}

\paragraph{C, problem 21 (header restoration).}
The CGS precondition mentions $\texttt{INT\_MIN}$ and $\texttt{INT\_MAX}$.
The first generated source omits \texttt{<limits.h>}, so Frama-C aborts during
specification parsing.  VGCR restores the include; both target clauses remain
covered and the program returns to joint success.

\paragraph{Rust, problem 23 (mutability repair).}
CGS adds the frame fact $\texttt{ensures *v == *old(v)}$, but the generated
signature uses an immutable reference where the implementation expects a
mutable one.  VGCR repairs the reference mutability and preserves the specification,
turning the type error into a verified artifact.  This is a code/interface
repair, not a specification weakening.

\subsection{Code generation performance is sensitive to received specification}
\label{subsec:sensitive}
\paragraph{Oracle specifications can remove failures that VGCR cannot repair.}
For C problem 10 (\texttt{array\_double.c}), the self-spec Direct artifact
times out with 12 of 13 goals proved, and self-spec VGCR does not recover it.
The failing proof interface omits the loop variable from the loop frame and
does not expose the arithmetic range needed for doubling array elements.  In
the oracle-conditioned Direct artifact, the loop frame includes both the index
and the modified array range, and the contract makes the no-overflow condition
explicit; Frama-C/WP proves all 14 goals.  Rust problem 42
(\texttt{VecCopyPrefix}) exhibits the same sensitivity at the contract-syntax
level:

\begin{minipage}{\linewidth}
\begin{lstlisting}[style=casecode]
Self:   ensures dst.len() == old(dst).len()
Oracle: ensures final(dst).len() == old(dst).len()
        ensures forall|i: int| 0 <= i < n ==> final(dst)[i] == src[i]
\end{lstlisting}
\end{minipage}

Verus rejects the self-generated form because a mutable-reference
postcondition must disambiguate the pre- and post-state with
\texttt{old}/\texttt{final}; the oracle-conditioned Direct implementation
verifies.  These cases show that a specification can hinder downstream
generation through its proof interface even when repair is available.

\paragraph{Oracle specifications can also be harder to satisfy.}
The direction is not universal.  For Rust problems 54
(\texttt{SquareBounded}) and 59 (\texttt{SafeMulSmall}), the self-spec Direct
implementations verify, whereas the oracle-conditioned implementations fail
with possible arithmetic overflow.  The oracle contracts require the model to
establish safe multiplication, an obligation absent from the self-generated
contracts.  Similarly, Java problem 16 (\texttt{ArrayMin}) verifies under the
self-generated Direct contract but fails under the oracle-conditioned artifact
because the generated \texttt{loop\_writes} clause omits modified locals.  An
oracle contract may therefore reveal real safety obligations while also making
the associated code-and-proof task strictly more demanding.

\paragraph{Equal validity can conceal a large adequacy gap.}
The clearest example is Kimi-K2.7-Code on the Rust track.  Direct generation
has exactly the same validity under Self and Oracle specifications, yet the
self-generated contracts cover substantially fewer requirement targets:

\begin{table}[h]
\centering
\caption{Kimi-K2.7-Code Direct results on Rust.  Oracle contracts have full
target coverage by construction, whereas equal validity under Self does not
imply adequate generated specifications.}
\label{tab:kimi-rust-oracle-case}
\small
\setlength{\tabcolsep}{9pt}
\begin{tabular}{lccc}
\toprule
Specification source & Validity & Coverage & Joint success \\
\midrule
Self   & 91 & 0.697 & 57 \\
Oracle & 91 & 1.000 & 91 \\
\bottomrule
\end{tabular}
\end{table}

At the problem level, the equal validity totals arise from offsetting changes.
Oracle contracts recover four failures (\texttt{OptionIncrement},
\texttt{IsEven}, \texttt{ResultMapIncrement}, and
\texttt{VecSwapFirstLast}) but make four previously valid cases fail
(\texttt{VecZeroPrefix}, \texttt{SquareBounded}, \texttt{SafeMulSmall}, and
\texttt{VecTruncateOne}).  The unchanged total of 91 therefore hides both
directions of specification sensitivity, while the 34-point joint-success gap
exposes the missing requirement coverage in the self-generated contracts.

\subsection{Takeaway}

Across these artifacts, CGS improves the semantic completeness of specifications
and therefore micro coverage, but can lower validity when it introduces alias,
quantifier, disjunction, or exceptional-proof obligations.  VGCR is useful
precisely at this boundary: it uses verifier diagnostics to change the
implementation while keeping the contract frozen.  The resulting behavior
explains why the aggregate table can show higher coverage and higher final
joint success even when CGS alone lowers verifier acceptance.  The oracle
cases further show that verifier acceptance and specification adequacy are
non-interchangeable: validity is sensitive to the supplied proof interface,
while joint success additionally detects contracts that omit required behavior.

\section{Prompt Records}
\label{app:prompts}

This appendix gives compact, semantically complete records of every prompt
stage used by the released pipeline, with runtime fields represented by their
names (requirement, specification, code, and diagnostics).  System prompts require
either strict JSON or code only, while user prompts define the schema and the
frozen-specification boundary.  The executable templates contain the same
instructions with literal JSON schemas and language-specific placeholders.
CGS means the constraint-extraction, constraint-to-specification, and
self-check/refinement prompts.  VGCR means the repair-analysis and focused
candidate prompts.  The same stage order is used in all four native verifier
tracks.

\subsection{C / ACSL}

\begin{lstlisting}[style=prompt]
SYSTEM (specification): You are an expert in ACSL specification design for C code.
Return strict JSON only.

USER (specification): Given the requirement below, design a minimal but useful
ACSL function specification. Return JSON with function_signature, acsl_block,
code_annotation_hints (loop_invariants, loop_assigns, loop_variants), and notes.
The signature hint must be copied exactly. The ACSL block is a function
contract only; put loop annotations in code_annotation_hints.

SYSTEM (CGS extraction): You are an expert in translating requirements into
verification constraints. Return strict JSON only.

USER (CGS extraction): Extract structured verification constraints from the
requirement. Return function_signature, atomic preconditions, postconditions,
invariants, and notes. Keep every constraint testable and concise.

SYSTEM (CGS translation): You are an expert ACSL specification engineer.
Convert constraints into a complete ACSL contract. Return strict JSON only.

USER (CGS translation): Create ACSL spec JSON using requirement + structured
constraints + signature hint. Include requires/assigns/ensures; keep loop
annotations outside the function contract and use valid/valid_read for pointers.

SYSTEM (CGS self-check): You are a strict requirement-spec alignment reviewer.
Return strict JSON only.

USER (CGS self-check): Check whether the generated specification misses
requirement constraints. Return is_aligned, missing_constraints,
inconsistent_items, refinement_hints, and notes. Use empty arrays when aligned.

SYSTEM (CGS refinement): You refine ACSL specs to improve requirement
alignment. Return strict JSON only.

USER (CGS refinement): Refine the generated specification from the original
requirement, structured constraints, current specification, and alignment
findings. Return the same specification schema. Preserve the signature; keep
the ACSL block as a function contract and move loop annotations to code hints.

SYSTEM (code generation): You are an expert C developer writing
verification-friendly code. Output C code only.

USER (code generation): Implement one C function from the requirement,
signature, frozen ACSL block, and code-annotation hints. Keep exactly one
matching implementation, keep the contract directly above it, do not change
the contract, and place loop annotations immediately before their loops.

SYSTEM (VGCR analysis): You are a verification-guided C/ACSL repair planner.
Return strict JSON only.

USER (VGCR analysis): Analyze a failed Frama-C/WP attempt. Return a failure
summary, typed subgoals with verifier evidence and repair hints, a global
strategy, and risk_notes. Treat the ACSL function contract as frozen.

SYSTEM (VGCR candidate): You are an expert C developer using
prove-as-you-generate repair. Output C code only.

USER (VGCR candidate): Repair the implementation using the verifier-derived
plan. Keep the frozen ACSL block directly above the function; do not change,
weaken, reorder, or delete contract clauses. Change only the body and
statement-level loop annotations. Output complete C source and no explanation.
\end{lstlisting}

\subsection{Java / JML}

\begin{lstlisting}[style=prompt]
SYSTEM (specification): You are an expert in JML specification design for
Java code. Return strict JSON only.

USER (specification): Given the Java requirement, class name, signature hint,
and fixed type context, return function_signature, jml_block,
code_annotation_hints, helper_declarations, and notes. Use OpenJML-compatible
requires, assignable, ensures, and exceptional behavior. Preserve exact field
names and narrow mutation frames.

SYSTEM (CGS extraction): You are an expert in translating Java requirements
into verification constraints. Return strict JSON only.

USER (CGS extraction): Extract atomic preconditions, postconditions,
frame_conditions, exceptional_behaviors, invariants, and helper declarations.
Include nullability, bounds, frames, object invariants, overflow, and
exception constraints when relevant.

SYSTEM (CGS translation): You are an expert JML specification engineer.
Convert Java verification constraints into a complete OpenJML-compatible JML
contract. Return strict JSON only.

USER (CGS translation): Create JML spec JSON using requirement, constraints,
signature hint, and exact type context. Emit independent clauses, narrow
assignable frames, and valid \old/\result/\forall/\exists syntax.

SYSTEM (CGS self-check): You are a strict Java requirement/JML-spec alignment
reviewer. Return strict JSON only.

USER (CGS self-check): Check for missing or inconsistent Java constraints.
Reject renamed fields, widened frames, combined atomic requirements, and
conditional weakenings. Return is_aligned, missing_constraints,
inconsistent_items, refinement_hints, and notes.

SYSTEM (CGS refinement): You refine JML specs to improve Java requirement
alignment. Return strict JSON only.

USER (CGS refinement): Refine the current JML specification using the original
requirement, structured constraints, alignment findings, and fixed type
context. Preserve exact field names, atomic clauses, narrow frames, and
unconditional guarantees; keep loop annotations outside the method contract.

SYSTEM (code generation): You are an expert Java developer writing
OpenJML-friendly code. Output Java source code only.

USER (code generation): Implement one complete source file from the
requirement and frozen JML specification. Define one public class with the
required name, keep the contract directly above the method, include required
helpers, and add only implementation-level loop annotations.

SYSTEM (VGCR analysis): You are a verification-guided Java/JML repair planner.
Return strict JSON only.

USER (VGCR analysis): Analyze the OpenJML result and produce typed subgoals,
evidence, repair hints, a global strategy, and risk notes. Treat the JML method
contract as frozen; prefer body, helper, and loop-annotation changes.

SYSTEM (VGCR candidate): You are an expert Java/OpenJML developer using
prove-as-you-generate repair. Output Java source code only.

USER (VGCR candidate): Repair the implementation from the verifier-derived
plan. Keep the frozen JML contract directly above the method and do not change,
weaken, reorder, or delete clauses. Output one complete public class.
\end{lstlisting}

\subsection{Rust / Verus}

\begin{lstlisting}[style=prompt]
SYSTEM (specification): You are an expert in Rust verification with Verus.
Return strict JSON only.

USER (specification): Given the Rust requirement and signature hint, return
function_signature, verus_contract, atomic verus_clauses, code_annotation_hints,
and notes. Use named returns, old/final for mutable references, and v@ sequence
views. Do not invent null, validity, ownership, wf, or panics predicates.

SYSTEM (CGS extraction): You are an expert in translating Rust requirements
into Verus verification constraints. Return strict JSON only.

USER (CGS extraction): Extract atomic preconditions, postconditions,
panic_freedom, mutation_frame, and invariants. Include ownership/borrowing,
vector bounds, Option/Result cases, overflow, and mutation effects when needed.

SYSTEM (CGS translation): You are an expert Verus specification engineer.
Convert Rust constraints into a complete Verus function contract. Return strict
JSON only.

USER (CGS translation): Emit requires and ensures for every functional case,
return value, mutation, and unchanged region. Use only v@.len(), v@[i as int],
old(v)@, final(v)@, and valid Verus datatype projections.

SYSTEM (CGS self-check): You are a strict Rust requirement/Verus-spec alignment
reviewer. Return strict JSON only.

USER (CGS self-check): Check functional coverage and reject invented sequence
members, dropped return cases, or dropped mutation/frame constraints. Return
is_aligned, missing_constraints, inconsistent_items, refinement_hints, and notes.

SYSTEM (CGS refinement): You refine Verus specs to improve Rust requirement
alignment. Return strict JSON only.

USER (CGS refinement): Refine the Verus specification from the requirement,
constraints, current specification, and findings. Preserve every correct
functional clause; add or correct missing cases without deleting return,
mutation, or unchanged-region facts. Keep loop annotations in code hints.

SYSTEM (code generation): You are an expert Rust developer writing
Verus-friendly code. Output Rust source code only.

USER (code generation): Implement one complete Rust/Verus file from the
requirement and frozen contract. Include vstd::prelude, one verus! block, and
main outside it. Preserve requires/ensures semantics, use safe Rust, and add
implementation-level invariants or decreases clauses when needed.

SYSTEM (VGCR analysis): You are a verification-guided Rust/Verus repair
planner. Return strict JSON only.

USER (VGCR analysis): Analyze the Verus result and return typed subgoals,
verifier evidence, concrete repair hints, a global strategy, and risk notes.
Treat the Verus contract as frozen.

SYSTEM (VGCR candidate): You are an expert Rust/Verus developer using
prove-as-you-generate repair. Output Rust source code only.

USER (VGCR candidate): Repair the implementation from the plan while keeping
the frozen requires/ensures semantics. Change only body, helper proof code,
and loop annotations; use safe Rust and preserve the verus! wrapper and main.
\end{lstlisting}

\subsection{Python / Nagini}

\begin{lstlisting}[style=prompt]
SYSTEM (specification): You are an expert in Python formal verification with
Nagini. Return strict JSON only.

USER (specification): Given the Python requirement and signature hint, return
function_signature, nagini_contract, atomic nagini_clauses,
code_annotation_hints, and notes. Use Requires, Ensures, Acc, Old, Result(),
Implies, list_pred, and dict_pred as appropriate; keep the contract first in
the function body.

SYSTEM (CGS extraction): You are an expert in translating Python requirements
into Nagini verification constraints. Return strict JSON only.

USER (CGS extraction): Extract atomic preconditions, postconditions,
permission_conditions, exception_freedom, and invariants. Preserve boundary
cases and use Old(...) for values read from pre-state containers.

SYSTEM (CGS translation): You are an expert Nagini specification engineer.
Convert Python constraints into a complete Nagini contract. Return strict JSON
only.

USER (CGS translation): Emit atomic Requires/Ensures/Acc clauses. Use Result(),
Implies(condition, conclusion), list_pred/dict_pred, Old(...), and ordinary
Python Boolean expressions. Do not use wildcard, Forall, lambda triggers, or
unsupported helper APIs.

SYSTEM (CGS self-check): You are a strict Python requirement/Nagini-spec
alignment reviewer. Return strict JSON only.

USER (CGS self-check): Check the original requirement as well as extracted
constraints. Return is_aligned, missing_constraints, inconsistent_items,
refinement_hints, and notes; preserve mutation facts, None cases, permissions,
and exception freedom.

SYSTEM (CGS refinement): You refine Nagini specs to improve Python requirement
alignment. Return strict JSON only.

USER (CGS refinement): Refine the Nagini specification using the requirement,
constraints, current specification, and findings. Preserve correct atomic
clauses, mutation and permission facts, boundary cases, and legal Nagini
syntax; keep loop invariants in code hints.

SYSTEM (code generation): You are an expert Python developer writing
Nagini-friendly code. Output Python source code only.

USER (code generation): Implement one complete Python source file from the
requirement and frozen Nagini specification. Keep the target signature and
contract as the first function statements, add no tests or top-level execution,
and use simple Nagini-compatible code and supplied invariants only.

SYSTEM (VGCR analysis): You are a verification-guided Python/Nagini repair
planner. Return strict JSON only.

USER (VGCR analysis): Analyze Nagini diagnostics and produce typed subgoals,
evidence, repair hints, a global strategy, and risk notes. Treat the Nagini
contract as frozen and do not suggest unsupported Fold/Unfold or element APIs.

SYSTEM (VGCR candidate): You are an expert Python/Nagini developer using
prove-as-you-generate repair. Output Python source code only.

USER (VGCR candidate): Repair the implementation using the plan. Preserve the
target signature and frozen contract semantics; change only body logic and
Invariant annotations. Keep the required imports and use simple Nagini Python.
\end{lstlisting}

\end{document}